\documentclass[letter]{aa}  

\usepackage{graphicx}
\usepackage{txfonts}
\usepackage{lipsum}
\usepackage{subcaption}         
\usepackage{lscape}             
\usepackage{placeins}           

\usepackage[]{hyperref}
\hypersetup{colorlinks=true,citecolor=blue}
\usepackage{multirow}

\begin{document}

   \title{Investigating aerosols as a way to reconcile K2-18\,b JWST MIRI and NIRISS/NIRSpec observations}

   \titlerunning{Investigating aerosols on K2-18\,b}

   \author{A. Y. Jaziri\inst{1,2}
        \and T. Drant\inst{3}
        }

   \authorrunning{Jaziri and Drant}

   \institute{LATMOS/IPSL, UVSQ Universit\'{e} Paris-Saclay, Sorbonne Universit\'{e}, CNRS, Guyancourt, France
            \and Laboratoire d’astrophysique de Bordeaux, Univ. Bordeaux, CNRS, B18N, all\'{e}e Geoffroy Saint-Hilaire, 33615 Pessac, France
            \and 
            ETH University, Center for Origin and Prevalence of Life, Department of Earth and Planetary Sciences, 8092 Zurich, Switzerland.\\
            \email{yassin.jaziri@latmos.ipsl.fr}}

   \date{Received 19 August 2025; Accepted 09 October 2025}

  \abstract
   {Recent JWST observations of the temperate sub-Neptune K2-18 b with NIRISS SOSS/NIRSpec G395H and MIRI LRS have yielded apparently inconsistent results: the MIRI spectra exhibit spectral features nearly twice as large as those seen at shorter wavelengths, challenging the high-metallicity, CH$_4$-rich non-equilibrium model that fits the NIRISS/NIRSpec data. We perform a suite of atmospheric retrievals on both datasets, including free-chemistry, non-equilibrium, and aerosol models, using laboratory-derived complex refractive indices for a variety of photochemical haze analogues. Free retrievals systematically return lower metallicities than inferred by self-consistent chemical disequilibrium models, and the inclusion of absorbing aerosols, especially CH$_4$-dominated, nitrogen-poor tholins, can further reduce the inferred metallicity by over an order of magnitude. These hazes reproduce the observed NIRISS slope through scattering and match MIRI features via C–H bending absorption near 7 $\mu$m, while yielding particle properties consistent with photochemical production in H$_2$-rich atmospheres. Although their inclusion improves the joint fit and reduces tension between datasets, it also significantly lowers the retrieved CH$_4$ abundance, highlighting degeneracies between metallicity, composition, and aerosol properties. Our results underscore the importance of aerosol absorption in interpreting temperate sub-Neptune spectra, and motivate future JWST observations and laboratory work to break these degeneracies.}


   \maketitle

\nolinenumbers
\section{Introduction}

Recent observations with JWST NIRISS SOSS/NIRSpec G395H \citep{Madhusudhan2023} and MIRI LRS \citep{Madhusudhan2025} of the temperate sub-Neptune K2-18\,b have sparked several debates regarding molecular detections and atmospheric characterization. The detection of CH$_4$ and CO$_2$ using NIRISS SOSS/NIRSpec G395H was initially reported by \citet{Madhusudhan2023}, but the presence of CO$_2$ was later challenged and mitigated by \citet{Schmidt2025}, before being recently confirmed by \citet{Hu2025}. A comprehensive non-equilibrium study confirmed the detection of CH$_4$, while also suggesting a high atmospheric metallicity ($\sim$266) and an elevated C/O ratio ($\ge$2.1) \citep{Jaziri2025}.

Subsequently, the MIRI LRS observations \citep{Madhusudhan2025} reignited the discussion on molecular detections. However, these data yielded only marginal detection significance \citep{Welbanks2025,Taylor2025}. A new reduction of the MIRI LRS data, combined with NIRISS SOSS/NIRSpec G395H observations \citep{Luque2025}, reached similar conclusions: a robust detection of CH$_4$, while other molecular signatures remained statistically insignificant.

Nevertheless, one striking aspect of the MIRI LRS observations is the consistently higher amplitude of spectral features compared to those seen in the NIRISS SOSS/NIRSpec G395H data, regardless of the reduction method. These features, which formed the basis of claimed molecular detections in \citet{Madhusudhan2025}, appear inconsistent across the full set of observations. This can be seen in the joint analysis by \citet{Luque2025}.

Given the apparent slope at shorter wavelengths in the NIRISS SOSS data and the high inferred abundance of methane \citep{Madhusudhan2023,Madhusudhan2025,Hu2025}, aerosols are likely present in the atmosphere of K2-18\,b, such as methane clouds or photochemical hazes. Scattering-induced extinction by these atmospheric particles is known to flatten the spectra at shorter wavelengths without significantly affecting the longer wavelengths. They were for instance proposed to explain the flat spectra of the temperate sub-Neptune GJ 1214 b \citep{Gao2023}. The presence of photochemical hazes was also predicted in the simulations of \cite{Wogan2024} for K2-18 b. The absorption features observed in the K2-18 b MIRI LRS observations may be attributed to these hazes, reflecting the C-H bending resonances of the solid particles. In this work, we explore the role of aerosols in an effort to reconcile the various JWST observations of K2-18\,b. All retrieval data can be downloaded from \texttt{Zenodo}\footnote{\url{https://doi.org/10.5281/zenodo.16277833}}.

\section{Methods}
\label{sect: method}

Using the MIRI LRS data reduced with \texttt{Eureka!} \citep{Bell2022}, from \citet{Luque2025}, we performed atmospheric retrievals with \texttt{TauREx 3}\footnote{\url{https://github.com/ucl-exoplanets/taurex3}} \citep{Al-Refaie2021}, including hydrocarbons as motivated by the non-equilibrium models of \citet{Jaziri2025}. In combination with the NIRISS SOSS/NIRSpec G395H analysis from \citet{Jaziri2025}, this approach aims to emphasize the discrepancies between the two datasets. While the results are data- and model-dependent, we provide here a consistent analysis with one retrieval code of a given set of reduced data, including in particular a complementary analysis of the MIRI LRS data reduced with \texttt{Eureka!}, which was not detailed in \citet{Luque2025}.

We started by retrieving a flat spectrum to assess the statistical significance of any spectral features, for both dataset and combined dataset. Subsequently, we carried out retrievals including CH$_4$, CO$_2$, CO, H$_2$O, NH$_3$, C$_2$H$_2$, C$_2$H$_4$, H$_2$CO, H$_2$S, SO$_2$, and HCN, which allowed us to identify the key absorbers for the combined analysis. We included these molecules because they have consistent absorption cross-sections computed by ExoMol in hydrogen-rich atmospheres \citep{Chubb2021}. We acknowledge that our reduced sample of hydrocarbons may introduce some quantitative bias in the results. However, \citet{Welbanks2025} showed that many hydrocarbons contribute with similar significance, and that the most abundant species (C$_2$H$_2$, C$_2$H$_4$, and C$_2$H$_6$) likely have comparable abundances (see also \citealp{Jaziri2025}). For this reason, we prioritized consistency in the opacity data.

We also carried out a retrieval including all molecules using the NIRISS SOSS/NIRSpec G395H data reduced with \texttt{JExoRES} \citep{Madhusudhan2023} to derive a metallicity that is directly comparable through the same methodology. Additionally, we performed a non-equilibrium chemistry retrieval using \texttt{TauREx 3} coupled with FRECKLL \citep{alrefaie2024}. While this approach remains computationally expensive and not yet fully stable, it was used to compare our results with non-equilibrium chemistry findings from the NIRISS SOSS/NIRSpec G395H data \citep{Jaziri2025} and the consistency of the MIRI LRS observations under non-equilibrium conditions. These results are described in appendix \ref{ap: miri_results}.

Additionally to the gray cloud parametrization used in all retrievals, we then used \texttt{TauREx 3} together with the \texttt{TauREx-PyMieScatt}\footnote{\url{https://github.com/groningen-exoatmospheres/taurex-pymiescatt}} module \citep{Changeat2025} to retrieve the combined dataset, including aerosol contributions, and adopting a free-chemistry approach with a subset of relevant molecules selected based on retrievals performed without aerosols. An offset of –41 ppm between the NIRISS SOSS and NIRSpec G395H datasets was adopted from \citet{Madhusudhan2023}. For the MIRI LRS dataset, an offset of +160 ppm was applied, determined by fitting it to the best retrieved non-equilibrium model from \citet{Jaziri2025} for the NIRISS SOSS/NIRSpec G395H data (see details in Appendix~\ref{ap: offset}).

The contribution of aerosols is accounted for using Mie scattering calculations. Previous strategies relied on the use of a simplified function to include aerosol scattering throughout the Rayleigh and Mie regimes (e.g., \citealp{Kitzmann2018}). This approach ignores the absorbing properties of the aerosols, encoded in the imaginary part of the complex refractive index, since absorption is usually negligible compared to scattering for small particles in the Visible and near-IR spectral range. In the present work, we include the absorbing properties of photochemical hazes that could explain the features observed in the K2-18 b observations at longer wavelengths in addition to the scattering effect at shorter wavelengths in the NIRISS SOSS/NIRSpec G395H range. The haze extinction is thus calculated using complex refractive index data obtained on laboratory haze analogues called tholins. The absorption properties of photochemical hazes are known to vary with the gas composition, temperature, pressure and irradiation efficiency \citep{Brasse2015,Sciamma2023,Drant2024}. We thus used 7 different datasets compiled from the original study of \citet{Khare1984} and the recent data of \citet{Drant2025} to explore the influence of the haze absorbing properties on the retrieval and final posterior distribution. We performed simulations with the refractive index data of exoplanet haze analogs that were produced using two different experimental setups (PAMPRE at LATMOS and COSmIC at NASA Ames) using only CH$_4$ as a reactive gas. In addition, we performed simulations with the refractive index data obtained on Titan tholins using various experimental setups and N$_2$/CH$_4$ ratio. The different refractive index data are summarized in Table \ref{tab: haze}. In the simulations, haze particles are assumed spherical with a lognormal size distribution, making the particle radius the only microphysical parameter in the retrieval. Photochemical hazes can form fractal aggregates, which increase the effective radius and scattering efficiency. Accounting for aggregates would require additional poorly constrained parameters, such as the number of monomers per particle and the fractal dimension, increasing degeneracy in the retrievals (e.g., \citealp{Adams2019}). Therefore, only spherical particles were considered.

\begin{table}[h]
 \centering
 \caption[]{\label{tab: haze}Summary of the different haze refractive index data considered}
\begin{tabular}{lccc}
 \hline \hline
Tholins      & Gas composition  & Experimental setup    \\ \hline
\noalign{\smallskip}
exo1     & 95 \% Ar - 5\% CH$_4$ & PAMPRE            \\
exo2     & 95 \% Ar - 5\% CH$_4$ & COSmIC            \\
Titan1     & 90 \% N$_2$ - 10\% CH$_4$ & PAMPRE            \\
Titan2     & 90 \% N$_2$ - 10\% CH$_4$ & COSmIC            \\
Titan3     & 95 \% N$_2$ - 5\% CH$_4$ & PAMPRE            \\
Titan4     & 95 \% N$_2$ - 5\% CH$_4$ & COSmIC            \\
Titan5     & 90 \% N$_2$ - 10\% CH$_4$ & \cite{Khare1984}            \\
\hline
\end{tabular}
\end{table}

See the priors and retrieval setup in Appendix~\ref{ap: taurexret}. We use the Bayes Factor (B), as defined in \citet{Benneke2013}, to assess the statistical significance of one model compared to another. This metric allows us to compare two models using their log-evidences (ln(Z)) and evaluate the relative significance of one retrieval compared to another. A value of ln(B) between 1 and 2.5 indicates weak preference, between 2.5 and 5 indicates moderate preference, and values above 5 indicate strong preference for one model over the other. However, these thresholds should be interpreted with caution, as recent work by \citet{Kipping2025} has shown that the strength of evidence may be less decisive than previously assumed.


\section{Reconciling NIRISS SOSS/NIRSpec G395H and MIRI LRS observations with aerosols}
\label{sect: joint_discussion}

Previous studies have retrieved high abundances of CH$_4$, ranging from 1\% to over 10\% \citep{Madhusudhan2023, Schmidt2025, Jaziri2025, Hu2025}. Such significant methane levels suggest the formation of aerosols, such as Titan-like hazes via photochemistry (e.g., \citealp{Trainer2006} and \citealp{Arney2016}). Considering the modeled temperatures from previous studies \citep{Charnay2021, Blain2021}, along with our retrieved low temperatures, the formation of methane clouds is also plausible. These clouds may exhibit similar opacities to photochemical hazes generated in H$_2$-dominated atmospheres. Both hazes and clouds could contribute to the flattening of the spectra at shorter wavelengths while preserving strong features at longer wavelengths. This could explain the differences in feature amplitudes observed between NIRISS SOSS/NIRSpec G395H and MIRI LRS observations.

In addition, photochemical hazes poor in nitrogen, as expected in H$_2$-dominated sub-Neptune atmospheres, present significant absorption features attributed to C-H stretching (3.4 - 3.7 $\mu$m) and bending (6.7 - 7.2 $\mu$m) resonances \citep{Drant2025} which partly overlap with the bands of gaseous methane and thus could be also one of the gases consistent with the K2-18 b observations. Similar absorption bands would also be expected for methane clouds making them also consistent with the observed absorption features.

The combined analysis is driven by the NIRISS SOSS/NIRSpec G395H data, which have already been confirmed to contain significant spectral features, most notably CH$_4$, contrary to the MIRI LRS data. This is further supported by the retrieved metallicity, which remains consistent with values obtained using only NIRISS SOSS/NIRSpec G395H data (approximately 40 with a free-chemistry model).

We also observe that the combined analysis does not improve the CO$_2$ detection, which is expected since CO$_2$ has no noticeable feature in the mid-IR, and shows even less support for C$_2$H$_4$ compared to retrievals based solely on MIRI LRS data. The retrieved abundance of CH$_4$ is also consistent with previous data reduction analysis from \citet{Luque2025} (see Figure \ref{fig: corner_plot_free}).

\begin{table}[h]
 \centering
 \caption[]{\label{tab: nirret}Free retrieval results for combined observations}
\begin{tabular}{lccc}
 \hline \hline
Model                    & ln(Z)    & ln(B)     & metallicity               \\ \hline
\noalign{\smallskip}
Flat line                & 28999.08 & Reference & -                         \\
Worst model              & 29007.74 & 8.66      & 21.07$^{+18.48}_{-16.92}$ \\ \hline
\noalign{\smallskip}
All molecules            & 29006.25 & Reference & 40.85$^{+8.49}_{-12.87}$  \\
\noalign{\smallskip}
CH$_4$/C$_2$H$_4$/CO$_2$ & 29009.54 & 3.29      & 19.63$^{+14.53}_{-13.57}$ \\ \hline
\noalign{\smallskip}
CH$_4$/C$_2$H$_4$/CO$_2$ & 29009.54 & Reference & 19.63$^{+14.53}_{-13.57}$ \\
\noalign{\smallskip}
With Tholins exo1        & 29015.07 & 5.53      & 0.29$^{+1.27}_{-0.22}$    \\
\noalign{\smallskip}
No CO$_2$                & 29007.74 & 1.84      & 21.07$^{+18.48}_{-16.92}$ \\
\noalign{\smallskip}
With Tholins exo2        & 29011.21 & 1.67      & 1.53$^{+28.05}_{-1.46}$   \\
\noalign{\smallskip}
With Tholins Titan1      & 29010.59 & 1.05      & 2.14$^{+30.89}_{-2.08}$   \\
\noalign{\smallskip}
With Tholins Titan5      & 29010.27 & 0.73      & 0.22$^{+24.04}_{-0.18}$   \\
\noalign{\smallskip}
With Tholins Titan2      & 29009.92 & 0.38      & 20.64$^{+15.51}_{-19.59}$ \\
\noalign{\smallskip}
With Tholins Titan3      & 29009.76 & 0.07      & 0.53$^{+24.95}_{-0.48}$   \\
\noalign{\smallskip}
With Tholins Titan4      & 29009.47 & -0.07     & 26.74$^{+13.20}_{-12.64}$ \\
\noalign{\smallskip}
No C$_2$H$_4$            & 29009.68 & -0.14     & 14.71$^{+17.30}_{-11.64}$ \\
\hline
\end{tabular}
\end{table}

\begin{figure*}[h!]
\centering
\includegraphics[width=\hsize]{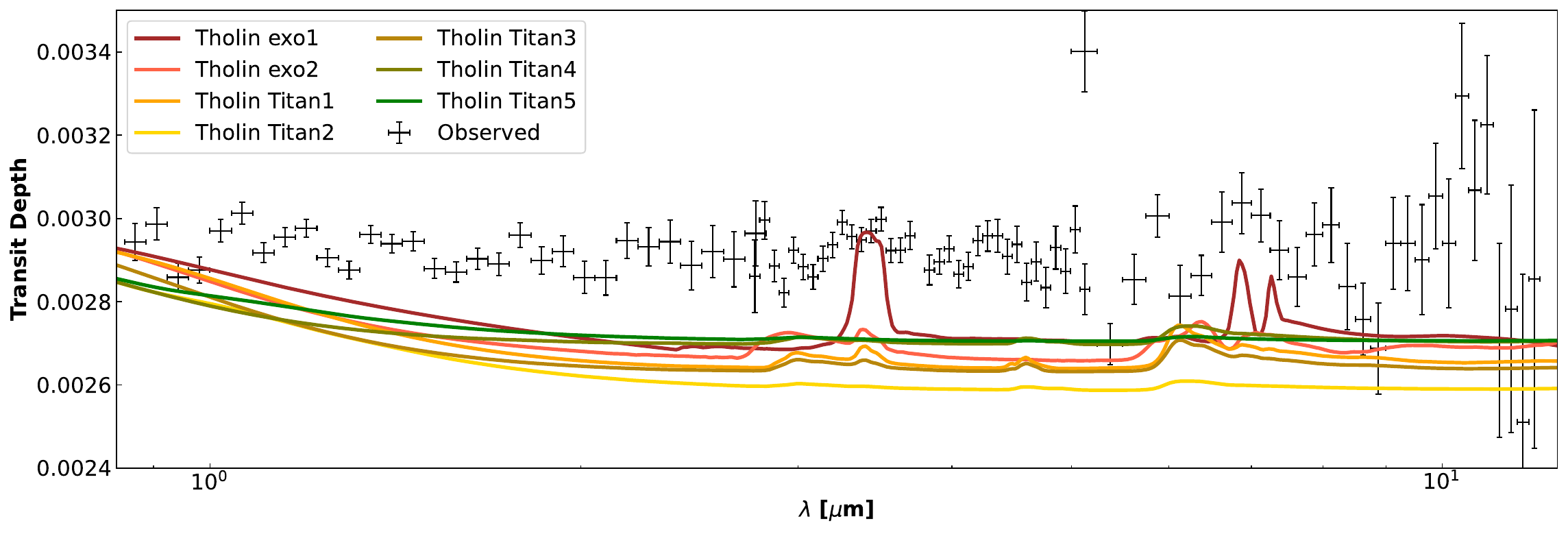}
  \caption{Combined spectra of NIRISS SOSS/NIRSpec G395H low resolution \citep{Madhusudhan2023} and MIRI LRS \texttt{Eureka!} data reduction \citep{Luque2025} (error bars) are shown alongside the retrieved tholins contributions (solid lines) of the different tholins data for each corresponding combine retrieval.}
     \label{fig: contrib_tholins}
\end{figure*}

For the tholins retrievals, the log-evidence values in Table \ref{tab: nirret} reveal discrepancies depending on the specific tholins data used. These discrepancies are explained by the variability in the strength and position of absorption features between the different datasets. One dataset in particular, tholins exo1, is strongly favored. Unlike the others, except for tholins exo2, tholins exo1 lacks features associated with nitrogen bonds and instead shows only –CH$_x$ signatures. The tholins exo2, although produced from a similar gas composition, exhibits C=C absorption features which are not observed in the K2-18 b observations. Only the C-H haze features of the tholins exo1 are consistent with the observed spectral features, especially the 7$\mu$m band previously attributed to DMS and DMDS by \citet{Madhusudhan2025}. The contributions of the various tholins datasets to the retrieved spectra are shown in Figure \ref{fig: contrib_tholins}, highlighting the spectral specificity of tholins exo1. The different Titan tholins datasets, including the one of \cite{Khare1984} used in the previous work of \cite{Wogan2024}, are not as favored as exo1 which is explained by the stronger absorbing power of these hazes produced in N$_2$-rich gas mixtures and the different N-H, C$\equiv$N, C=N absorption features which are not consistent with the K2-18 b observations. This analysis emphasizes the importance of the haze refractive index, especially the imaginary part of the refractive index that reflects absorption, since it influences the retrieval and can help rule out some haze composition based on the observed features at longer wavelengths. 

The best-fit model includes tholins exo1 and is preferred over models without tholins, though the preference is at the threshold between strong and moderate significance. The spectrum and its contributions are shown in Figure \ref{fig: combine_pec}, illustrating two key points:

\paragraph{Increased scale height:} Compared to retrievals without tholins, the atmosphere's base appears at a lower transit depth, decreasing from $\sim$0.0028 to $\sim$0.0025. This corresponds to a significantly lower retrieved metallicity, dropping from 19.63 to 0.29, relative to the reference retrieval (see Table \ref{tab: nirret}). This reflects the well-known degeneracy between cloud-top pressure and atmospheric metallicity. It also leads to a substantially lower retrieved CH$_4$ abundance, reduced by about two orders of magnitude (see Figure \ref{fig: corner_plot_free} and \ref{fig: corner_plot_tholins}).

\paragraph{Tholins spectral contribution:} Tholins exo1 contributes to the observed features across multiple spectral ranges: near 7$\mu$m in the MIRI LRS range and around 3.4$\mu$m in the NIRSpec G395H range (alongside CH$_4$) as absorption dominates, and at shorter wavelengths in the NIRISS SOSS range, affecting the spectral slope as scattering dominates. Notably, the retrieved CH$_4$ abundance decreases from 6.3×10$^{-2}$ to 3.0×10$^{-4}$ when considering these photochemical hazes compared to the reference model. The haze properties retrieved are also reasonable and in agreement with the modeling predictions of \cite{Lavvas2019} made for the temperature sub-Neptune GJ 1214 b. For the retrieval considering exo1 tholins, we constrain the particle radius in log unit to -1.65$^{+0.09}_{-0.12}$ (in $\mu$m), the haze density in log unit to 11.22$^{+0.36}_{-0.36}$ (m$^{-3}$), the mean pressure of the haze layer in log unit to -0.04$^{+1.02}_{-0.98}$ (in Pa), and the dP of the haze layer in log unit to 2.05$^{+1.06}_{-0.87}$ (in Pa), see Figure \ref{fig: corner_plot_tholins}. The retrievals considering hazes also tend to increase the atmospheric temperature (see Figure \ref{fig: temperature}) which along with the haze properties and the methane abundance affects the strength of the spectral features. Although this result reflects a degenerate behavior between different parameters influencing extinction, its physical meaning is reasonable as haze absorption in the UV-Visible tend to increase the atmospheric temperature of the upper atmosphere \citep{Arney2016}. This increased temperature driven by the haze heating properties could inhibit the formation of methane clouds and prevent the co-existence with the solid organic hazes.

While the spectral fit is broadly consistent when considering photochemical hazes, the resulting decrease in the retrieved CH$_4$ abundance goes against the initial motivation for considering their presence. The presence of photochemical hazes produced from gaseous methane is however generally consistent with the current data. 

\section{Conclusions}

We have investigated the apparent inconsistency between JWST NIRISS SOSS/NIRSpec G395H and MIRI LRS transmission spectra of the temperate sub-Neptune K2-18 b. The MIRI LRS observations exhibit apparent spectral features with amplitudes roughly twice those seen in the NIRISS SOSS/NIRSpec G395H data, which cannot be reproduced by the high-metallicity non-equilibrium model that fits the shorter-wavelength observations. Free-chemistry and non-equilibrium retrievals on the MIRI LRS spectrum instead favor lower metallicities and molecular compositions that are inconsistent with the NIRISS/NIRSpec results, yet none of the tested molecular models are statistically preferred over a flat spectrum for MIRI LRS alone.

To reconcile the datasets, we explored the role of aerosols, specifically photochemical hazes and methane clouds, which can flatten spectra at shorter wavelengths while preserving or enhancing features at longer wavelengths. Using laboratory-derived complex refractive indices for a range of haze analogues, we find that only hazes produced in CH$_4$-dominated mixtures without nitrogen and C=C features (tholins exo1) yield absorption signatures consistent with the combined NIRISS, NIRSpec, and MIRI data. These hazes naturally reproduce the observed slope in the NIRISS SOSS range through scattering, while contributing C–H bending absorption features near 7 $\mu$m in the MIRI LRS range.

The inclusion of such hazes improves the combined fit and moderately reduces the tension between the two datasets, though at the cost of significantly lowering the retrieved CH$_4$ abundance. The retrieved haze particle sizes, number densities, and layer pressures are broadly consistent with theoretical predictions for H$_2$-dominated sub-Neptune atmospheres, and the associated temperature increases in the upper atmosphere are physically plausible given aerosol absorption in the optical.

Overall, our results highlight the importance of aerosol absorption, particularly the imaginary component of the refractive index, in interpreting the atmospheres of temperate sub-Neptunes. While photochemical hazes offer a promising explanation for the discrepancy between short- and long-wavelength observations of K2-18 b, the current data do not yet allow a definitive determination of their composition or abundance. Future JWST observations and higher precision, alongside expanded laboratory measurements of exoplanet haze analogues, will be essential to break degeneracies between metallicity, molecular composition, and aerosol properties.


\begin{acknowledgements}
This project has received funding from the European Research Council (ERC) under the ERC OxyPlanets projects (grant agreement No. 101053033). This project acknowledges funding from the European Research Council (ERC) under the European Union's Horizon 2020 research and innovation programme (grant agreement No. 679030/WHIPLASH). T.D. acknowledges support by the 2024 NOMIS-ETH postdoc fellowship program. T.D. thanks the NOMIS foundation and ETH Zurich for funding via the 2024 Fellowship program. Some of the data presented in this paper were obtained from the Mikulski Archive for Space Telescopes (MAST) at the Space Telescope Science Institute. The specific observations analyzed can be accessed via \href{https://doi.org/10.17909/3ds1-8z15}{doi:10.17909/3ds1-8z15} and \href{https://doi.org/10.17909/rx29-yw62}{doi:10.17909/rx29-yw62}. We would like to thank the anonymous referee for his pertinent comments, which improved the presentation of our results.
\end{acknowledgements}

\bibliographystyle{aa}
\bibliography{biblio}


\begin{appendix}

\section{Dataset offset}
\label{ap: offset}

Using different telescope instruments can introduce biases between datasets, such as flux offsets. In this study, we work with three JWST datasets obtained from three different instruments: NIRISS SOSS, NIRSpec G395H, and MIRI LRS. For the NIRISS SOSS and NIRSpec G395H data, we use the observations from \citet{Madhusudhan2023}, along with their reported offset of –41 ppm. The MIRI LRS data are taken from the \texttt{Eureka!} reduction from \citet{Luque2025}, who found a flux offset ranging from approximately +120 ppm to +160 ppm depending on their retrieval priors or retrieval code.

\begin{figure}[h!]
\centering
\includegraphics[width=\hsize]{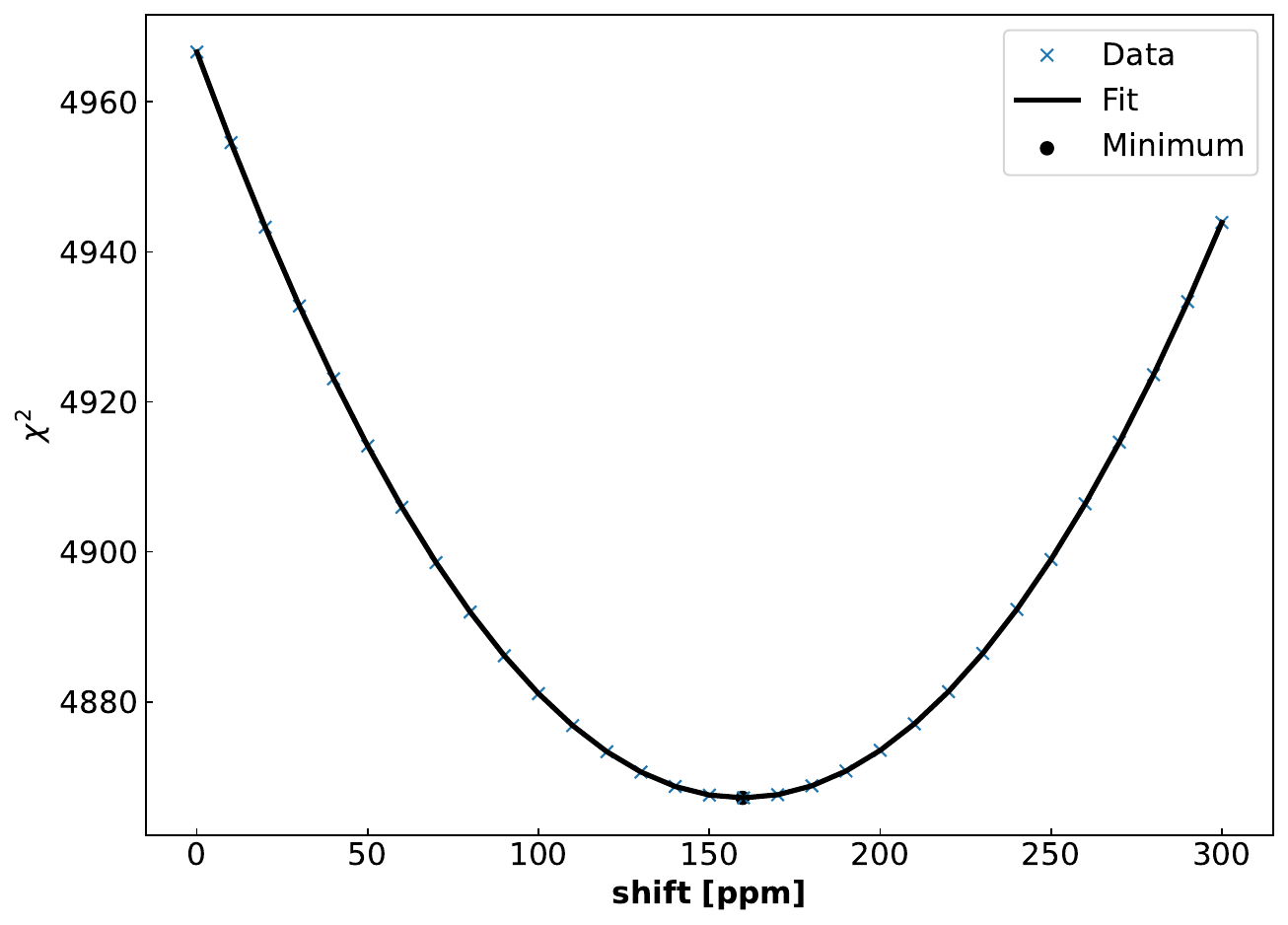}
  \caption{The $\chi^2$ of the best non-equilibrium model from \citet{Jaziri2025} is shown as a function of the MIRI LRS wavelength shift, based on the combined JWST observations from NIRISS SOSS, NIRSpec G395H \citep{Madhusudhan2023}, and MIRI LRS \citep{Luque2025}. Blue cross markers indicate the computed values, and a second-order polynomial fit is shown as a black solid line, with its minimum value (160 ppm) marked by a black dot.}
     \label{fig: shift}
\end{figure}

To ensure consistency with the non-equilibrium modeling results from \citet{Jaziri2025}, we use their best-fitting model to determine the optimal offset for the MIRI LRS dataset via a $\chi^2$ minimization approach. The resulting $\chi^2$ values as a function of the wavelength shift are shown in Figure~\ref{fig: shift}, along with a second-order polynomial fit. The minimum $\chi^2$ corresponds to an offset of +160 ppm, which we adopt for this study.

All these offsets are fixed to reduce the parameter space and accelerate aerosol retrievals. A more detailed investigation of their effects could be carried out in future work.

\section{Retrieval setup}
\label{ap: taurexret}

We performed four types of retrievals: flat-line, non-equilibrium chemistry, and constant-chemistry (with and without aerosols).

The flat-line retrieval was conducted by retrieving the planetary radius and a cloud deck, following the priors listed in Table \ref{tab: priors}. A minimal amount of H$_2$O (between 10$^{-12}$ and 10$^{-11}$) was included, as \texttt{TauREx 3} requires at least one absorber to run. The radius and the cloud deck are also retrieved for all other retrievals following the same priors.

For all other retrievals, the temperature profile was parameterized with 4-points. This is the minimum required to allow for a thermal inversion, capture deep-atmosphere temperatures relevant to disequilibrium chemistry and quenching processes, and assess aerosol impacts on the upper-atmosphere temperature. Temperature priors are detailed in Table \ref{tab: priors}.

The non-equilibrium chemistry retrieval was carried out using \texttt{FRECKLL} coupled with \texttt{TauREx 3}, without aerosols. K$_{zz}$ has been set to 10$^{10}$. Its value, within common range of values, has been shown in \citet{Jaziri2025} to not significantly affect the retrieval. The priors are detailed in Table \ref{tab: priors}. Due to its high computational cost and limited stability, we present only one non-equilibrium retrieval, applied to the MIRI LRS data, which contain relatively few data points. This allows us to assess the consistency of the MIRI LRS observations under non-equilibrium conditions. We used 800 live points for all retrievals.

We then used vertically constant profiles for the chemical species (commonly referred to as free chemistry), including all the molecules listed in Table \ref{tab: priors}, with their corresponding priors, for retrievals on the MIRI LRS data. This set of molecules represents the main species of interest that have been highlighted in the literature and are available in the ExoMol database, which is the most relevant source for opacities calculated for H$_2$-dominated atmospheres. Each molecule was removed individually to assess its significance compared to the reference case with all molecules. A subset of the most significant molecules was then selected to reduce computational cost in subsequent retrievals that include aerosols, which are more expensive to model. The aerosol priors are also provided in Table \ref{tab: priors}. This optimized subset of molecules was then used for retrievals of combined NIRISS SOSS, NIRSpec G395H, and MIRI LRS data, both with and without aerosols, using the same set of priors.

\begin{table}[!htbp]
    \centering
    \caption{Free parameters and priors for the retrievals.}
    \begin{tabular}{c|c}
    \hline
    \textbf{Parameters} & \textbf{Bounds} \\ \hline
    \multicolumn{2}{c}{\textbf{Common}} \\ \hline
    T$_{top}$ [K]                 & 50  to 700   \\
    T$_{bot}$ [K]                 & 300 to 1000  \\
    T$_{1}$ [K]                   & 50  to 700   \\ 
    T$_{2}$ [K]                   & 50  to 700   \\
    log$_{10}$(P$_{1}$) [Pa]      & 2   to 6     \\
    log$_{10}$(P$_{2}$) [Pa]      & -1  to 5     \\
    log$_{10}$(P$_{clouds}$) [Pa] & -2  to 6     \\
    radius [R$_{jup}$]            & 0.1 to 0.3   \\
    \hline
    \multicolumn{2}{c}{\textbf{Chemistry}} \\ \hline
    \textbf{Non-equilibrium}      &              \\
    log(Z)                        & -2 to 3      \\
    C/O                           & 0.01 to 10.0 \\
    \textbf{Constant}             &              \\
    log$_{10}$(H$_2$O)            & -12 to -1    \\
    log$_{10}$(CO)                & -12 to -1    \\
    log$_{10}$(CH$_4$)            & -12 to -0.15 \\
    log$_{10}$(CO$_2$)            & -12 to -1    \\
    log$_{10}$(HCN)               & -12 to -1    \\
    log$_{10}$(NH$_3$)            & -12 to -1    \\
    log$_{10}$(C$_2$H$_2$)        & -12 to -1    \\
    log$_{10}$(C$_2$H$_4$)        & -12 to -1    \\
    log$_{10}$(H$_2$CO)           & -12 to -1    \\
    log$_{10}$(SO$_2$)            & -12 to -1    \\
    log$_{10}$(H$_2$S)            & -12 to -1    \\
    \hline
    \multicolumn{2}{c}{\textbf{Aerosols}} \\ \hline
    log($\mu$) [$\mu$m]              & -2 to 1  \\
    log($\chi$) [molecules.m$^{-3}$] & 4  to 13 \\
    log(P) [Pa]                      & -2 to 5  \\
    log($\Delta$P) [Pa]              & -1 to 5  \\
    \end{tabular}
    \label{tab: priors}
\end{table}

\section{Inconsistency between NIRISS SOSS/NIRSpec G395H and MIRI LRS observations}
\label{ap: miri_results}

\begin{figure}[h!]
\centering
\includegraphics[width=\hsize]{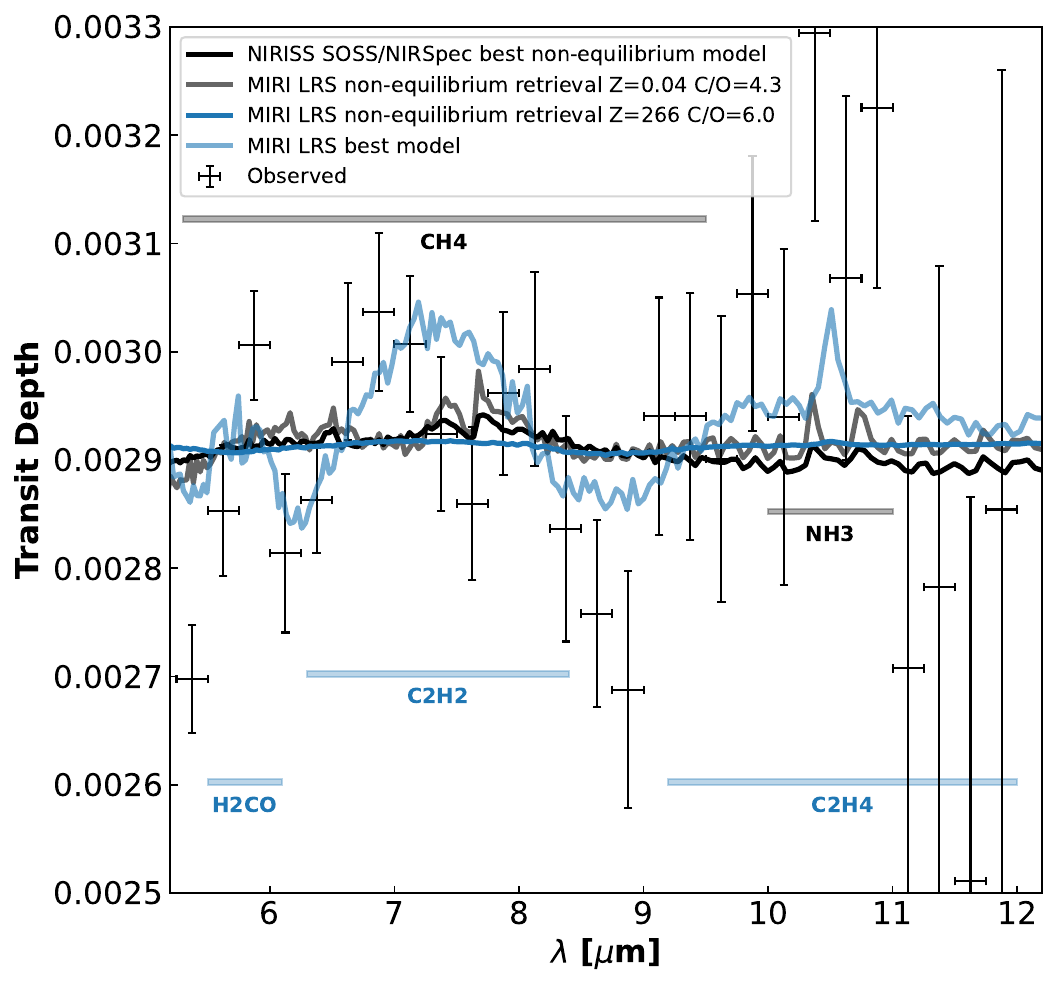}
  \caption{Spectrum from the MIRI LRS \texttt{Eureka!} data reduction \citep{Luque2025} (error bars) is compared to the best non-equilibrium model from \citet{Jaziri2025} (black solid line), which fits the NIRISS SOSS/NIRSpec G395H observations. They are also compared to the two MIRI LRS non-equilibrium retrieval solutions (light black solid line for low metallicity and blue solid line for high metallicity), as well as to the best MIRI LRS model (light blue solid line), all shown at a resolution of 200. Key spectral features are highlighted in their respective colors.}
     \label{fig: miri_pec}
\end{figure}

The NIRISS SOSS/NIRSpec G395H data exhibit feature amplitudes of approximately 200 ppm, whereas the MIRI LRS data show features at least twice as large. The relatively small amplitudes in the NIRISS SOSS/NIRSpec G395H data have been shown to favor high metallicity, with a value of 266 inferred using a non-equilibrium chemical model \citep{Jaziri2025}. However, the best-fit non-equilibrium model, using a grid of forward models with \texttt{FRECKLL}, from \citet{Jaziri2025} appears inconsistent with the larger spectral amplitudes seen in the MIRI LRS data, as illustrated in Figure \ref{fig: miri_pec}.

When applying a free-chemistry retrieval to the MIRI LRS data, we are able to fit the larger spectral features. The log-evidence results in Table \ref{tab: miriret} indicate that the three most favored molecules are C$_2$H$_2$, H$_2$CO, and C$_2$H$_4$. These molecules have absorption bands that align with the apparent features in the MIRI LRS spectrum, as shown in Figure \ref{fig: miri_pec}, where the best-fit model is overplotted on the observations. The retrieved metallicity is significantly lower, with a maximum value of 4.42$^{+14.31}_{-4.06}$, compared to the value of 266 reported by \citet{Jaziri2025}. It is important to note that \citet{Jaziri2025} used a non-equilibrium chemical model.

A retrieval using all molecules on the NIRISS SOSS/NIRSpec G395H data yields a metallicity of 40.56$^{+9.62}_{-13.35}$, demonstrating that, in our case, free-chemistry retrievals tend to bias metallicities toward lower values compared to non-equilibrium models. Following \citet{Welbanks2025}, this bias arises, at least in part, because we excluded from the retrieval certain species that may be abundant and influence the mean molecular weight but do not exhibit prominent spectral features, such as N$_2$, in order to improve computational efficiency. This effect is also seen in our results when comparing retrievals with the full set of molecules to those with the reduced set (see metallicities in Tables \ref{tab: miriret} and \ref{tab: nirret}). Nevertheless, the MIRI LRS retrievals, performed with the same setup as the NIRISS SOSS/NIRSpec G395H analysis and therefore directly comparable, tend to favor lower metallicity (from 40.56$^{+9.62}_{-13.35}$ to 4.42$^{+14.31}_{-4.06}$) to increase the scale height, allowing better fits to the larger observed features.

These results are inconsistent with those from the NIRISS SOSS/NIRSpec G395H analysis and the non-equilibrium chemical model, both of which favor higher metallicity and a high CH$_4$ abundance, as seen in Figure \ref{fig: miri_pec}. In contrast, the MIRI LRS retrievals indicate low metallicity and do not show a preference for CH$_4$. However, as shown in Table \ref{tab: miriret}, none of the tested models are statistically favored over a flat line. At best, they fall within the "weak preference" range. H$_2$CO is unlikely to be dominant. Its retrieved abundance of ~10$^{-5}$ exceeds the 3$\sigma$ upper limit derived by \citet{Jaziri2025}. C$_2$H$_2$ and C$_2$H$_4$ could plausibly be present, especially considering high C/O ratio and high metallicity. However, the retrievals favor C$_2$H$_2$ over CH$_4$, which is inconsistent with the NIRISS SOSS/NIRSpec G395H data. Our conclusions are consistent with previous analyses of MIRI LRS data using alternative reduction pipelines \citep{Madhusudhan2025, Taylor2025, Welbanks2025}, which likewise report no significant spectral signatures. A possible exception is a weak indication of H$_2$CO, though this is not statistically robust in a Bayesian framework and may simply reflect reduction-dependent systematics. In particular, features around 5-6 $\mu$m, where H$_2$CO absorption could occur, appear sensitive to the choice of data reduction, limiting the reliability of this tentative signal.

For the combined analysis, we therefore adopted CH$_4$, C$_2$H$_4$, and CO$_2$ as the key molecules. CO$_2$ remains somewhat uncertain in the NIRISS SOSS/NIRSpec G395H data, but was included for completeness since it has been recently confirmed with the new data from \citet{Hu2025}.

\begin{figure*}[h!]
\centering
\includegraphics[width=0.55\hsize]{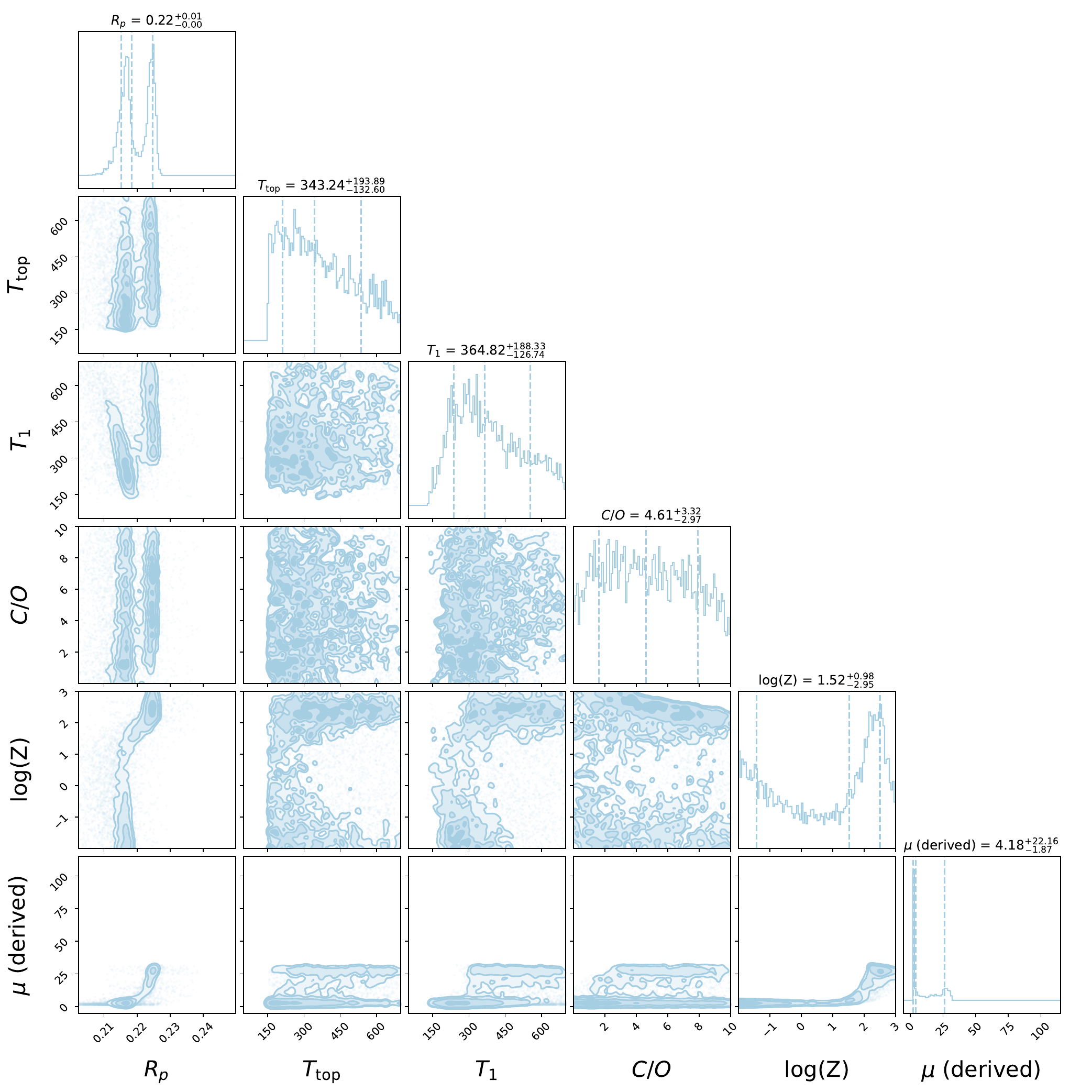}
  \caption{Corner plot of the non-equilibrium retrieval model with \texttt{FRECKLL} on the MIRI LRS \texttt{Eureka!} data reduction \citep{Luque2025}.}
     \label{fig: corner_plot_freckll}
\end{figure*}

Finally, a non-equilibrium chemistry retrieval on the MIRI LRS data reveals that the feature amplitudes are inconsistent with such models. The solution shows bimodality (see Figure \ref{fig: corner_plot_freckll}), with one mode corresponding to a low radius (0.2173$^{+0.0007}_{-0.0006}$ R$_{Jup}$) and low metallicity (0.04$^{+0.35}_{-0.02}$) to produce a high scale height, and the other corresponding to a higher radius (0.2245$^{+0.0006}_{-0.0009}$ R$_{Jup}$) and high metallicity (266$^{+243}_{-125}$). In the high-metallicity case, a higher C/O ratio (> 3.5) is also preferred, which enhances C$_2$H$_2$ and C$_2$H$_4$ features, similar to the free retrieval, but makes the result inconsistent with the NIRISS SOSS and NIRSpec G395H data that favor CH$_4$ and NH$_3$ features (see Figure \ref{fig: miri_pec}). Higher temperatures (> 340 K) are also favored. The feature amplitudes appear smaller compared to the data, although this mismatch is somewhat reduced in the low-metallicity scenario where CH$_4$ is dominant. Nonetheless, the feature amplitudes remain smaller than those predicted by the free-chemistry retrieval and the observed amplitudes (see Figure \ref{fig: miri_pec}).

Since none of the models are statistically significant, we cannot draw definitive conclusions, but these inconsistencies suggest that a key process may be missing from our current models.

\begin{table}[h!]
 \centering
 \caption[]{\label{tab: miriret}Retrieval results for MIRI LRS}
\begin{tabular}{lccc}
 \hline \hline
Model         & ln(Z)  & ln(B)     & metallicity \\ \hline
\noalign{\smallskip}
Flat line     & 185.11 & Reference & -           \\
Best model    & 187.56 & 2.45      & 0.28$^{+4.72}_{-0.27}$  \\ \hline
\noalign{\smallskip}
All molecules & 186.93 & Reference & 4.41$^{+13.98}_{-4.02}$ \\
\noalign{\smallskip}
No C$_2$H$_2$ & 185.90 & 1.03      & 4.42$^{+14.31}_{-4.06}$ \\
\noalign{\smallskip}
No H$_2$CO    & 186.08 & 0.85      & 4.28$^{+15.60}_{-3.96}$ \\
\noalign{\smallskip}
No C$_2$H$_4$ & 186.36 & 0.57      & 2.43$^{+12.66}_{-2.28}$ \\
\noalign{\smallskip}
No HCN        & 186.75 & 0.18      & 2.82$^{+12.92}_{-2.59}$ \\
\noalign{\smallskip}
No CO         & 186.78 & 0.15      & 4.14$^{+14.61}_{-3.80}$ \\
\noalign{\smallskip}
No CH$_4$     & 186.85 & 0.08      & 2.73$^{+12.41}_{-2.52}$ \\
\noalign{\smallskip}
No CO$_2$     & 186.86 & 0.07      & 3.20$^{+12.14}_{-2.94}$ \\
\noalign{\smallskip}
No NH$_3$     & 186.90 & 0.03      & 3.21$^{+13.83}_{-2.99}$ \\
\noalign{\smallskip}
No H$_2$O     & 186.91 & 0.02      & 3.55$^{+14.42}_{-3.24}$ \\
\noalign{\smallskip}
No H$_2$S     & 186.91 & 0.02      & 3.50$^{+12.58}_{-3.18}$ \\
\noalign{\smallskip}
No SO$_2$     & 187.21 & -0.28     & 3.41$^{+13.34}_{-3.10}$ \\ \hline
\noalign{\smallskip}
CH$_4$/C$_2$H$_2$/C$_2$H$_4$ & \multirow{2}{*}{187.26} & \multirow{2}{*}{Reference} & \multirow{2}{*}{0.62$^{+6.66}_{-0.59}$} \\
H$_2$O/H$_2$CO & & & \\
\noalign{\smallskip}
No C$_2$H$_2$ & 186.15 & 1.11      & 0.33$^{+6.24}_{-0.32}$ \\
\noalign{\smallskip}
No H$_2$CO    & 186.51 & 0.75      & 0.34$^{+5.54}_{-0.33}$ \\
\noalign{\smallskip}
No C$_2$H$_4$ & 186.51 & 0.75      & 0.04$^{+1.23}_{-0.03}$ \\
\noalign{\smallskip}
No H$_2$O     & 187.24 & 0.02      & 0.19$^{+3.47}_{-0.18}$ \\
\noalign{\smallskip}
No CH$_4$     & 187.56 & -0.30     & 0.28$^{+4.72}_{-0.27}$ \\
\noalign{\smallskip}
With Tholins exo1  & 186.83 & -0.43     & 1.26$^{+7.82}_{-1.19}$ \\
\hline
\end{tabular}
\end{table}

\onecolumn

\section{Joint NIRISS SOSS/NIRSpec G395H and MIRI LRS data additional materials}
\label{ap: joint_results}

\begin{figure*}[h!]
\centering
\includegraphics[width=\hsize]{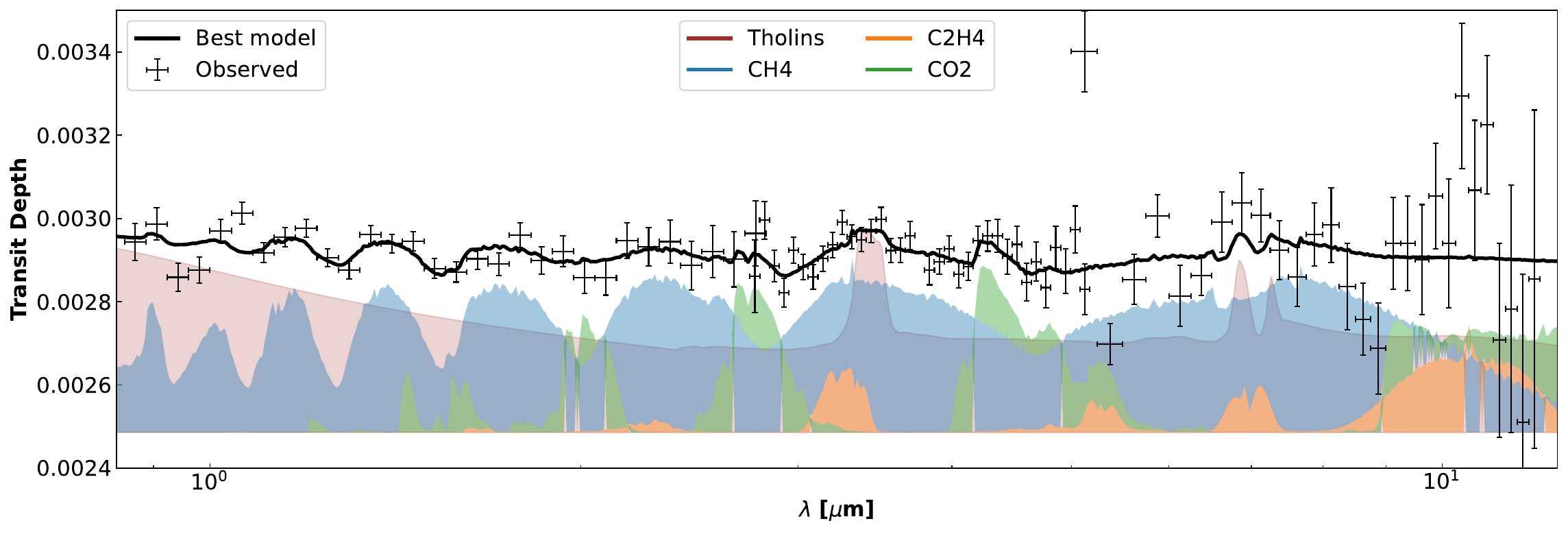}
  \caption{Combined spectra of NIRISS SOSS/NIRSpec G395H low resolution \citep{Madhusudhan2023} and MIRI LRS \texttt{Eureka!} data reduction \citep{Luque2025} (error bars) are shown alongside the best model (black solid line) and its contributions. It correspond to the retrieval including tholins exo1.}
     \label{fig: combine_pec}
\end{figure*}

\begin{figure}[h!]
\centering
\includegraphics[width=0.5\hsize]{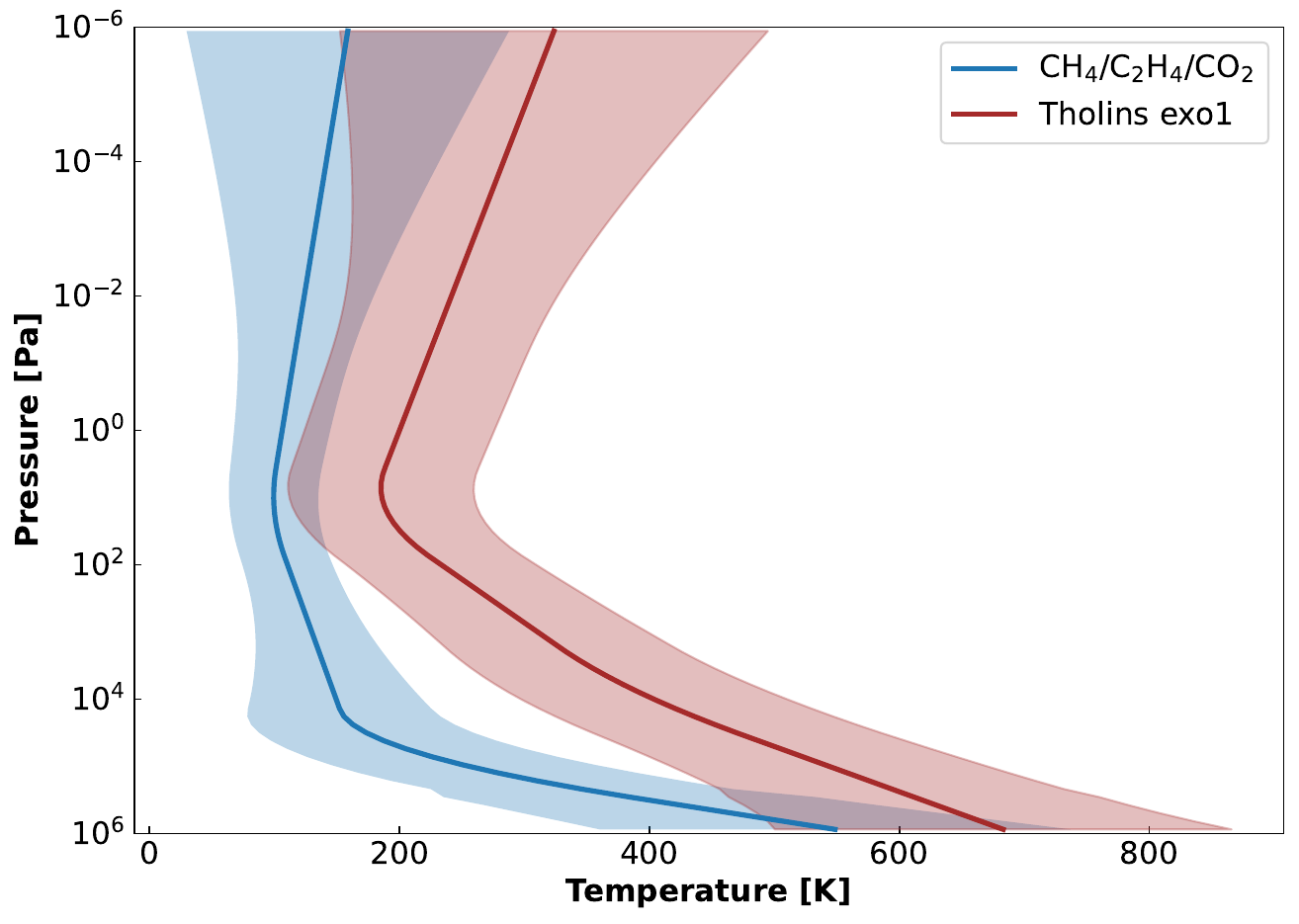}
  \caption{Retrieved temperature profile with its uncertainty from the retrieval using CH$_4$/C$_2$H$_4$/CO$_2$ (blue) and with the addition of tholins exo1 (brown), based on the combined observations.}
     \label{fig: temperature}
\end{figure}

\begin{figure*}[h!]
\centering
\includegraphics[width=\hsize]{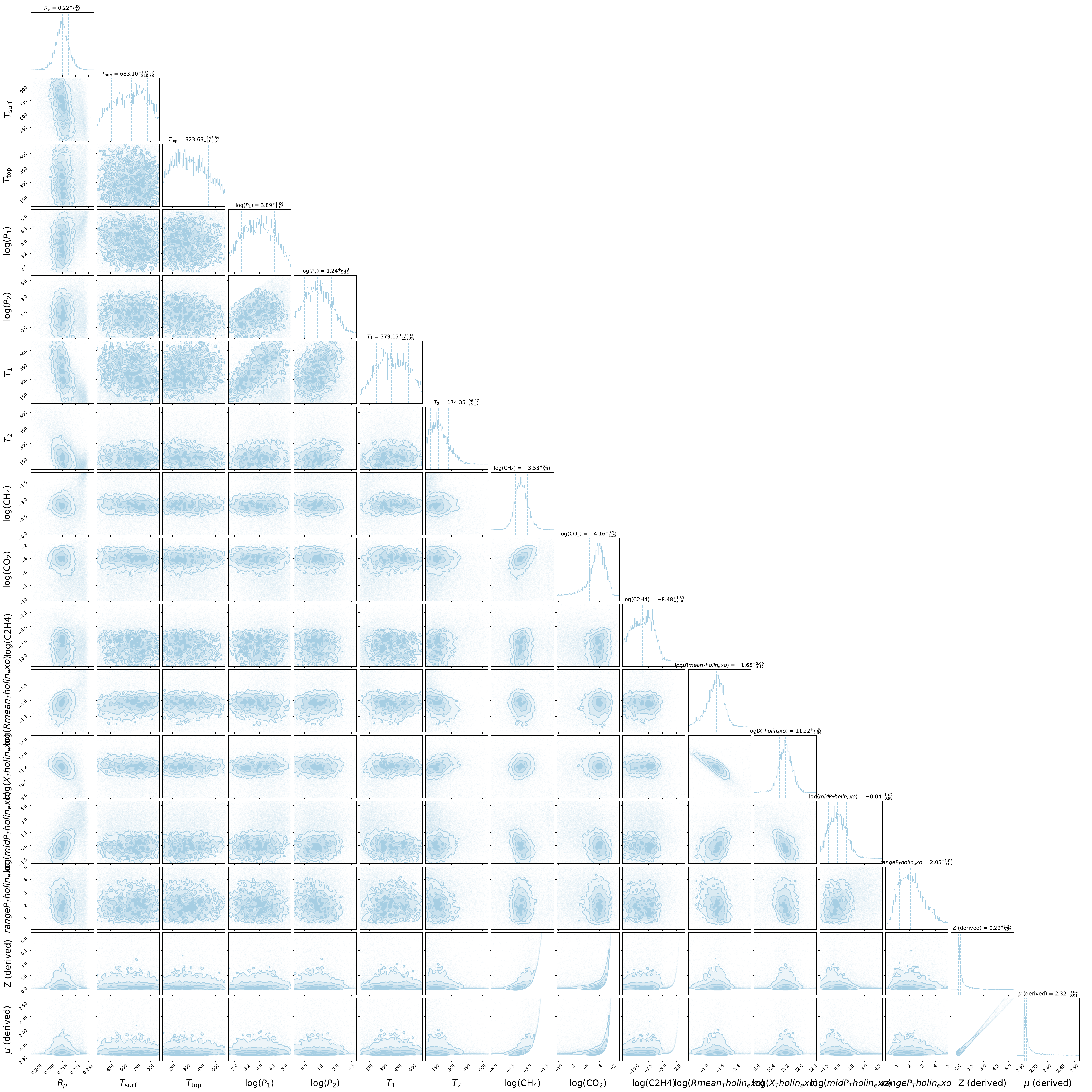}
  \caption{Corner plot of the best retrieval model on the combine spectra of NIRISS SOSS/NIRSpec G395H \citep{Madhusudhan2023} and MIRI LRS \texttt{Eureka!} data reduction \citep{Luque2025}. It correspond to the retrieval including tholins exo1.}
     \label{fig: corner_plot_tholins}
\end{figure*}

\begin{figure*}[h!]
\centering
\includegraphics[width=\hsize]{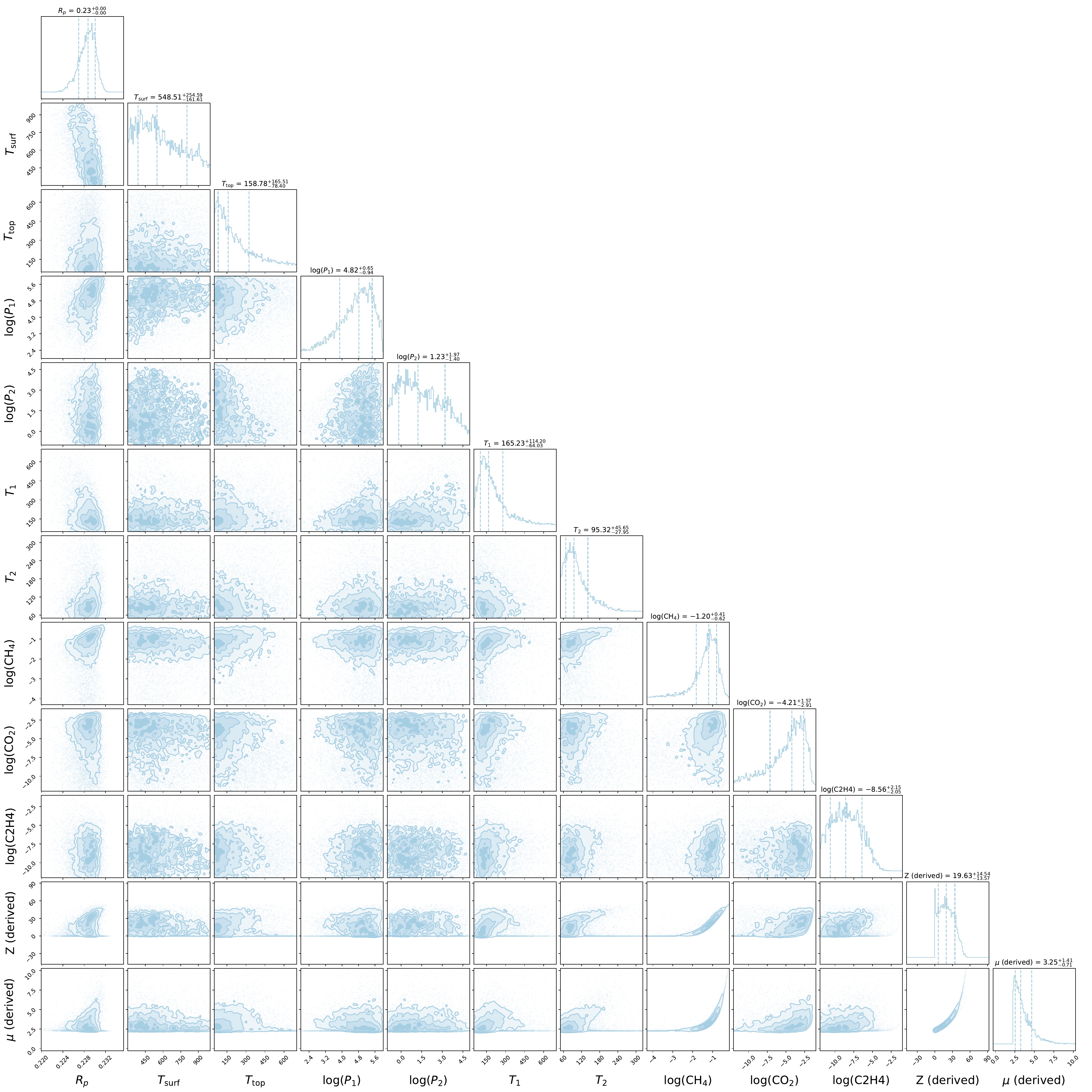}
  \caption{Corner plot of the free retrieval model with CH$_4$/C$_2$H$_4$/CO$_2$ on the combine spectra of NIRISS SOSS/NIRSpec G395H \citep{Madhusudhan2023} and MIRI LRS \texttt{Eureka!} data reduction \citep{Luque2025}.}
     \label{fig: corner_plot_free}
\end{figure*}

In this appendix, we present Figure \ref{fig: combine_pec} and Figure \ref{fig: corner_plot_tholins} respectively the spectral contributions and the corner plot of the best-fit model from the joint NIRISS SOSS/NIRSpec G395H and MIRI LRS retrieval, as well as the temperature profiles from two retrievals Figure \ref{fig: temperature}, with and without haze, to illustrate the effect of haze on the retrieved temperature, as discussed in Section \ref{sect: joint_discussion}. Figure \ref{fig: corner_plot_free} show the corner plot of free retrieval model with CH$_4$/C$_2$H$_4$/CO$_2$ on the combine spectra. Figure \ref{fig: corner_plot_free} show the corner plot of free retrieval model with CH$_4$/C$_2$H$_4$/CO$_2$ on the combine spectra. Our retrieved temperatures in Figure \ref{fig: temperature} are generally higher than the condensation curve for methane (see Fig 1 of \cite{Gao2021} for example) which may suggest that photochemical hazes are a more likely aerosol candidate than methane clouds to reconcile the K2-18 b JWST observations.

\end{appendix}
\end{document}